\pdfoutput=1
\documentclass[letter]{aa}  
\usepackage{graphicx,float}
\usepackage{latexsym,amsmath,amssymb}
\usepackage{natbib}
\usepackage[scriptsize,tight]{subfigure}
\bibpunct{(}{)}{;}{a}{}{,}
\usepackage[english]{babel}

\usepackage{textcomp}

\def \isdc        {INTEGRAL Science Data Centre, Universit\'e de Gen\`eve, Chemin d'Ecogia 16, CH-1290
                  Versoix, Switzerland}

\def \obsgen      {Observatoire de Gen\`eve, Universit\'e de Gen\`eve, Chemin des Maillettes 51, CH-1290
                Sauverny, Switzerland}
                
\def \iaat		  {Institut f$\mathrm{\ddot{u}}$r Astronomie und Astrophysik, Universit$\mathrm{\ddot{a}}$t T$\mathrm{\ddot{u}}$bingen, Sand 1, 72076 T$\mathrm{\ddot{u}}$bingen, Germany}

\begin{document}

\title{Twelve-hour spikes from the Crab Pevatron}


\author{M. Balbo\inst{1, 2}, R. Walter\inst{1, 2}, C. Ferrigno\inst{1, 2}, P. Bordas\inst{1, 3}}

\authorrunning{M. Balbo et al.}

\offprints{Matteo.Balbo@unige.ch}

\institute{\isdc \and \obsgen \and \iaat}

\date{Received October 22, 2010; refereed December 9, 2010; accepted December 15, 2010}
 
\abstract
{}
{The \object{Crab} nebula displayed a large $\gamma$-ray flare on September 18, 2010. To more closely understand the origin of this phenomenon, we analyze the  INTEGRAL (20-500\,keV) and FERMI (0.1-300\,GeV) data collected almost simultaneously during the flare.}
{We divide the available data into three different sets, corresponding to the pre-flare period, the flare, and the subsequent quiescence. For each period, we perform timing and spectral analyses to differentiate between the contributions of the pulsar and from the surrounding nebula to the $\gamma$-ray luminosity.}
{No significant variations in the pulse profile and spectral characteristics are detected in the hard X-ray domain. In contrast, we identify three separate enhancements in the $\gamma$-ray flux lasting for about 12 hours and separated by an interval of about two days from each other. 
The spectral analysis shows that the flux enhancement, confined below $\sim$1\,GeV, can be modelled by a power-law with a high energy exponential cut-off, where either the cut-off energy or the model normalization increased by a factor of $\sim5$ relative to the pre-flare emission. We also confirm that the $\gamma$-ray flare is not pulsed. }
{The timing and spectral analysis indicate that the $\gamma$-ray flare is due to synchrotron emission from a very compact Pevatron located in the region of interaction between the pulsar wind and the surrounding nebula. These are the highest electron energies ever measured in a cosmic accelerator.
The spectral properties of the flare are interpreted in the framework of a relativistically moving emitter and/or a harder emitting electron population.}

\keywords{Acceleration of particles -- Astroparticle physics -- Magnetic fields -- (Stars:) pulsars: individual: Crab -- Gamma rays: stars -- X-rays: stars}

\maketitle
%

\section{Introduction}

With an integrated luminosity of about $5\times10^{38}\,\mathrm{erg\,s^{-1}}$ and a distance of $\sim$2\,kpc, the \object{Crab} supernova remnant is very bright from the radio domain to TeV energies \citep[see e.g.,][for a review]{2008ARA&A..46..127H}. It is powered by a pulsar spinning on its axis in about 33\,ms that injects energetic electrons into the surrounding nebula. Nearly all the nebular emission up to 0.4\,GeV is believed to be produced by synchrotron cooling of these electrons in an average magnetic field of $\sim300\,\mu$G. At higher energies, inverse Compton (IC) cooling dominates.

The integrated high-energy flux of the nebula and the pulsar has been remarkably stable over the past few decades and the object is indeed used as a calibration source in several experiments \citep[but see ][ for the first report of a secular X-ray trend]{2010arXiv1010.2679W}. On 2010 September 22 the AGILE collaboration \citep{2010ATel.2855....1T} reported the first $\gamma$-ray flare from a source positionally consistent with the \object{Crab}, during which the flux above 100\,MeV was nearly double its normal value. The $\gamma-$ray flare from the direction of the \object{Crab} was confirmed by the Fermi collaboration \citep{2010ATel.2861....1B}.

During a period partially covering the $\gamma$-ray flare, the \object{Crab} region was observed by the INTEGRAL satellite and the Swift/BAT telescope during its routine sky survey. 
No statistically significant increase in the \object{Crab} flux
was observed in the IBIS/ISGRI light curves between 20 and 400 keV, as well as in the Swift/BAT $15-50\,$keV flux at a level of 5\% at the 1$\sigma$ confidence level \citep{2010ATel.2856....1F,2010ATel.2858....1M}.

Swift/XRT observed the \object{Crab} region for 1\,ks on 2010 September 22 at 16:42 UT and did not reveal any significant variation in the source flux, spectrum, and pulse profile \citep{2010ATel.2866....1E}. The ultraviolet and soft X-ray images excluded the presence of any bright unknown field object that could have contributed to the $\gamma$-ray flux \citep{2010ATel.2868....1H}. The near-infrared \object{Crab} flux in the J and H bands was also constant  \citep{2010ATel.2867....1K}. Dedicated pointing performed by RXTE did not show any significant change in the overall spectral properties in the 3-20\,keV band \citep{2010ATel.2872....1S}.

A 5ks TOO Chandra observation was performed on 2010 September 28 to monitor the morphology of the inner nebula. There are no particular variations with respect to the over 35 previous observations, with the possible exception of an anomalous extension of a bright knot closer (down to 3'') to the pulsar, also noticed in one archival observation \citep{2010ATel.2882....1T}. A subsequent HST optical observation of the \object{Crab} confirmed an increase in the emission about 3 arcsec south-east of the pulsar with respect to archival observations \citep{2010ATel.2903....1C}.  However, it remains unclear whether this feature is related to the $\gamma$-ray event.
The \object{Crab} $\gamma$-ray flux returned to its usual level on 2010 September 23, less than a week after the onset of the flare \citep{2010ATel.2879....1H}.


\section{Results}
\subsection{INTEGRAL} 
The INTEGRAL satellite is equipped with several high-energy instruments covering the energy range from a few keV to a few MeV. Here, we concentrate on the results from the spectrometer SPI \citep{spi} and the hard X-ray imager IBIS/ISGRI \citep{ibis,isgri}. We divided the available observations into two periods: the \emph{pre-flare}  from 2010 September 12 10:32 UT to 2010 September 18 05:56 UT (exposures: ISGRI 280\,ks, SPI 336\,ks), and the \emph{flare} from 2010 September 18 11:24 UT to 2010 September 23 10:20 UT (exposures: ISGRI 89\,ks, SPI 109\,ks). The gap between the two data sets is due to the uncertainty in the exact time of the flare onset. The data are analysed using the off-line science analysis software version 9 provided by the ISDC \citep{isdc}.  The data analysis is performed using the \texttt{SPIROS} package and the \texttt{GEDSAT} template background model, based on the count rate of the events saturating the Germanium detector.
The spectra obtained for both data sets are equal within the errors. For the pre-flare set, a best-fit to the data is obtained using a broken power-law model with a fixed energy break at 100 keV \citep{2009ApJ...704...17J}. The spectral indices are $\Gamma_{1} = 2.142 \pm 0.019$ and $\Gamma_{2} = 2.194 \pm 0.021$ ($\chi^{2} =$~42/47). For the data-set corresponding to the flare, a broken power-law model is also adequate, whose best-fit indices are $\Gamma_{1} = 2.103 \pm 0.038$ and $\Gamma_{2} = 2.149 \pm 0.041$ ($\chi^{2}$ = 52/47). The total flux between 50 and 1000 keV is, respectively, $2.79^{+0.09}_{-0.12} \times 10^{-8}$~erg~cm$^{-2}$~s$^{-1}$ before the flare and $ 2.81^{+0.14}_{-0.26} \times 10^{-8}$~erg~cm$^{-2}$~s$^{-1}$ during the flare, which are equal within the uncertainties. 

While the absolute count rate in IBIS/ISGRI is subject to a few percent variation due to instrumental systematic effects, the background-subtracted pulse profile normalized to its average count rate can be considered a trustful estimator of the pulse shape. We can thus compare the relative contribution of the pulsar and nebular emissions before and during the $\gamma$-ray flare. We exploited the method of \citet{2007ESASP.622..633S} and used the ephemeris of \cite{1993MNRAS.265.1003L}\footnote{\texttt{http://www.jb.man.ac.uk/$\sim$pulsar/crab.html}} to extract background-subtracted pulse profiles, accumulated in 100 phase bins in the energy ranges 20-40\,keV and 40-80\,keV, and in 25 bins in the 80-150 and 150-500\,keV energy ranges. 

\begin{figure}[t]
\resizebox{\hsize}{!}{\includegraphics{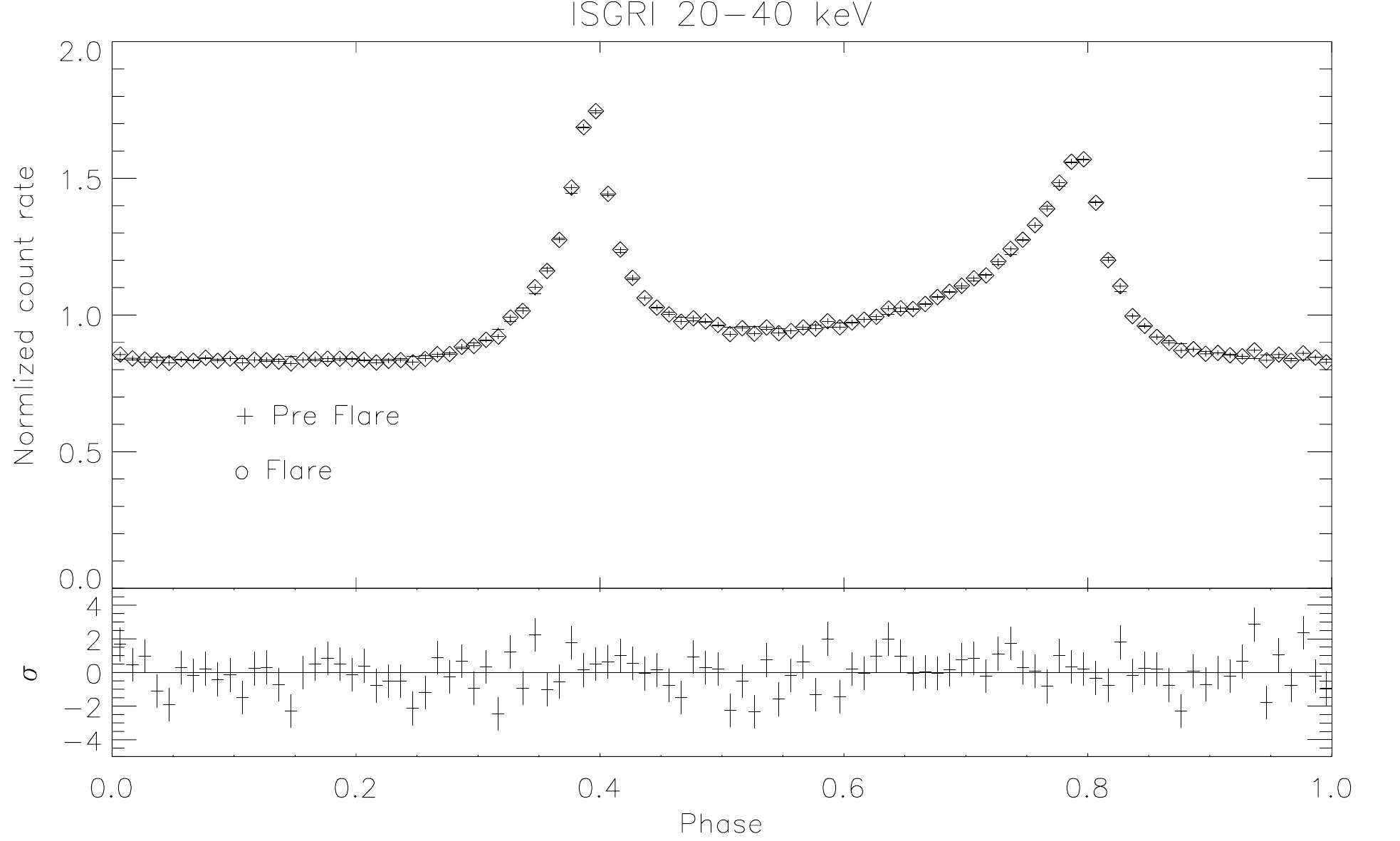}}
\caption{{\it Upper panel:} pulse profiles measured by IBIS/ISGRI in the 20-40\,keV energy band: crosses and diamonds represent the pulse profile before and during the \object{Crab} flare, respectively. {\it Lower panel:} significance in standard deviations of the difference between the pulse profile during and before the flare.}
\label{fig:pulse_isgri}
\end{figure}

The sum of the squared differences between the pulse profiles' normalized rates divided by the corresponding variances is distributed as a $\chi^2$ with a number of considered phase bins minus one degree of freedom.
In all energy bands, we found that the normalized pulse profiles are equal within a statistical uncertainty of one standard deviation (see e.g.,  Fig.~\ref{fig:pulse_isgri}, where $\chi^2_\mathrm{red}=1.18$ for 99 d.o.f.). We can estimate our sensitivity measuring a deviation from the template pulse profile, based on the assumption that the flux increase is constant throughout the phase.  In the 20-40\,keV energy range, we would be able to appreciate an increase in the un-pulsed emission  as small as 4\%  at the 99\% confidence level; our sensitivity diminishing to 6\% in the 40-80\,keV band, and 10\% between 80 and 150\,keV. Above 150\,keV, the sensitivity is limited to 30\%.

\subsection{FERMI}
FERMI/LAT is a pair-conversion telescope that operates in the 30~MeV-300~GeV energy range with unprecedented sensitivity and resolution \citep{2009ApJ...697.1071A}. In our analyses, we used all the events in the 0.1--300~GeV energy range belonging to the ``diffuse'' class, but rejected photons with zenith angles larger than $105^{\circ}$ to avoid contamination from the Earth's bright $\gamma$-ray albedo. We limited our sample to the data taken between 2010 September 1 to October 6. 
The analysis was performed using the ScienceTools, provided by the Fermi collaboration (Version v9r15p2). To obtain the source flux, we used the \texttt{gtlike} tool, which exploits a maximum likelihood method \citep{1996ApJ...461..396M}. The instrument response function is \texttt{P6\_V3\_DIFFUSE} and the Galactic emission is reproduced using the model \texttt{gll\_iem\_v02.fit}\footnote{http://fermi.gsfc.nasa.gov/ssc/data/analysis/software/\\
/aux/gll\_iem\_v02.fit}. All sources listed in the Fermi-LAT first year catalogue \citep{2010arXiv1002.2280T} within $7^{\circ}$ of the \object{Crab} position and with a TS detection greater than 10 are taken into account in the likelihood analysis.
   \begin{figure}[h!]
   \resizebox{\hsize}{!}{\includegraphics[width=0.45\textwidth]{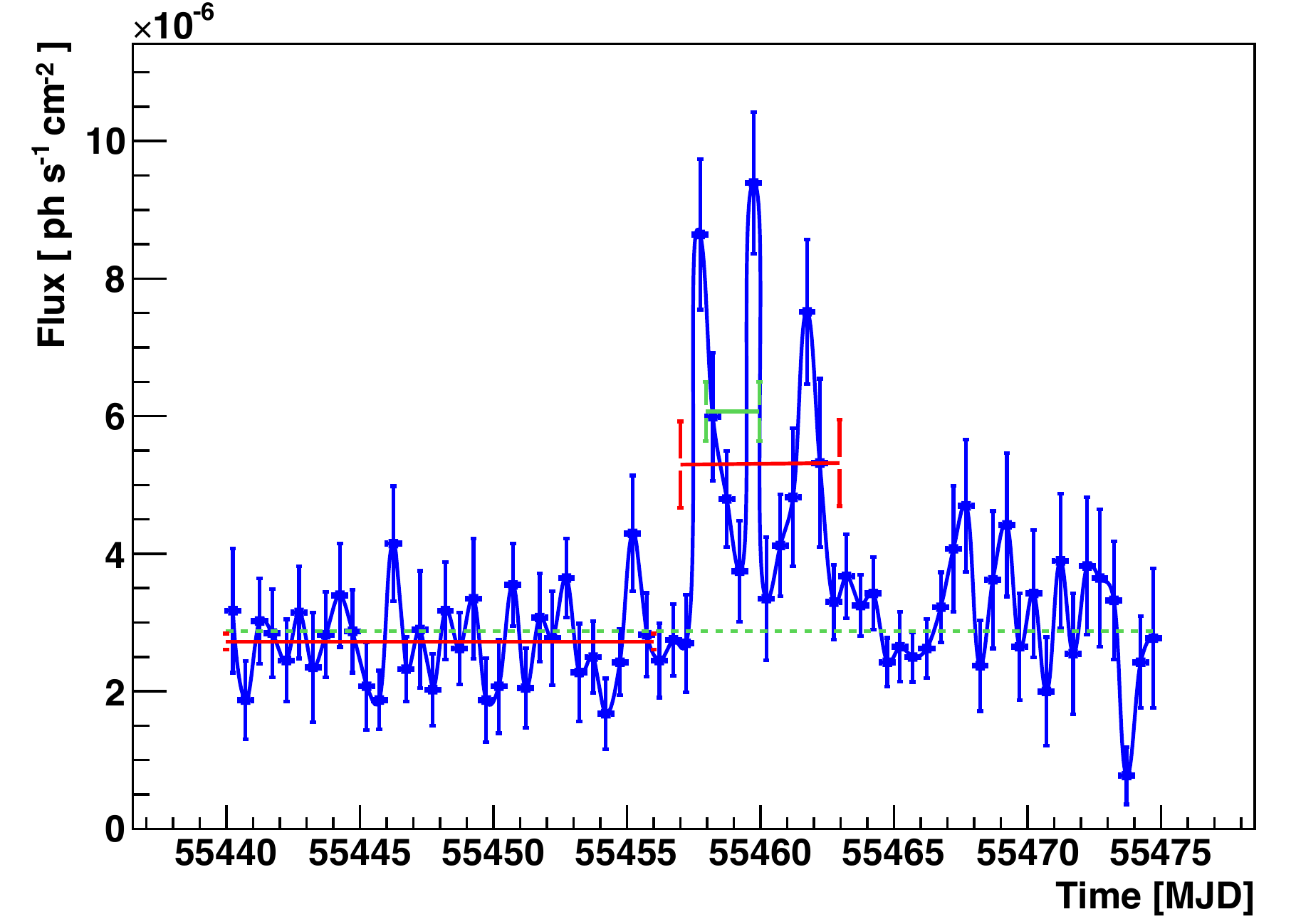}}
   \caption{Light curve of the \object{Crab} in the range 0.1--300 GeV with 12h time bins. The green dashed line indicates the flux of the \object{Crab} estimated over all the Fermi operation period. The green solid line is the average flux between September 19 and 21, as reported by \citet{2010ATel.2861....1B}. The red lines represent the average flux before and during the flares (see Table~\ref{tab:intervals}). The time is expressed in Modified Julian Date.}
   \label{fig:lc}
   \end{figure}
Each point in the source light-curve (Fig.~\ref{fig:lc}) is obtained by means of a single likelihood analysis. Owing to the low number of events collected in every time bin, we used a \texttt{PowerLaw2}\footnote{http://fermi.gsfc.nasa.gov/ssc/data/analysis/scitools/
/xml\_model\_defs.html\#powerlaw2} to represent the total emission of the \object{Crab}. To optimize the signal-to-noise ratio (S/N), we accumulated the light-curves with different binnings (6h, 12h, 1d, 2d), finding that the 12-hour binning is the most adequate to follow the \object{Crab} count-rate evolution. In Fig.~\ref{fig:lc}, the green dashed line indicates the average flux from the \object{Crab} of $(286 \pm 2)\times 10^{-8}~\mathrm{ph~cm^{-2}~s^{-1}}$, as reported by \cite{2010ATel.2861....1B}. Three distinct peaks are clearly visible, which exceed by a factor of nearly three the pre-flare flux level, corresponding to a detection significance $> 15\sigma$. The first steep increase occurs between 6:00 and 18:00 UTC on September 18. In 12 hours, the flare reaches a maximum flux of $(8.6 \pm 1.1)\times 10^{-6}~\mathrm{ph~cm^{-2}~s^{-1}}$. The second flux increase occurs about two days after, between 12:00 and 24:00 on September 20. The peak flux is $(9.4 \pm 1.0)\times 10^{-6}~\mathrm{ph~cm^{-2}~s^{-1}}$. The last flux increase was detected between 18:00 on September 21 and 18:00 on September 22, reaching a value of $(7.5 \pm 1.1)\times 10^{-6}~\mathrm{ph~cm^{-2}~s^{-1}}$. In all cases, the flare decay time was about 1 day. On the basis of the lightcurve reported in Fig.~\ref{fig:lc}, we divide the data into three different sets as reported in Table~\ref{tab:intervals}.

\begin{table}[h] 
\caption{Time intervals used to perform the Fermi analysis.}
\label{tab:intervals}
\begin{center}
\begin{tabular}{c | c | c | c}
\hline \hline \noalign{\smallskip} 
 & Pre-flare & Flare & Post-flare\\
\noalign{\smallskip\hrule\smallskip}
Start & 2010-09-01 & 2010-09-18 & 2010-09-26\\
$[UTC]$ & 00:00:00 & 00:00:00 & 00:00:00\\
\noalign{\smallskip\hrule\smallskip}
Stop & 2010-09-16 & 2010-09-23 & 2010-10-05\\
$[UTC]$ & 23:59:59 & 23:59:59 & 23:59:59\\
\noalign{\smallskip\hrule\smallskip}
\end{tabular}
\end{center}
\end{table}

   \begin{figure}[h!]
   \resizebox{\hsize}{!}{\includegraphics[width=0.45\textwidth]{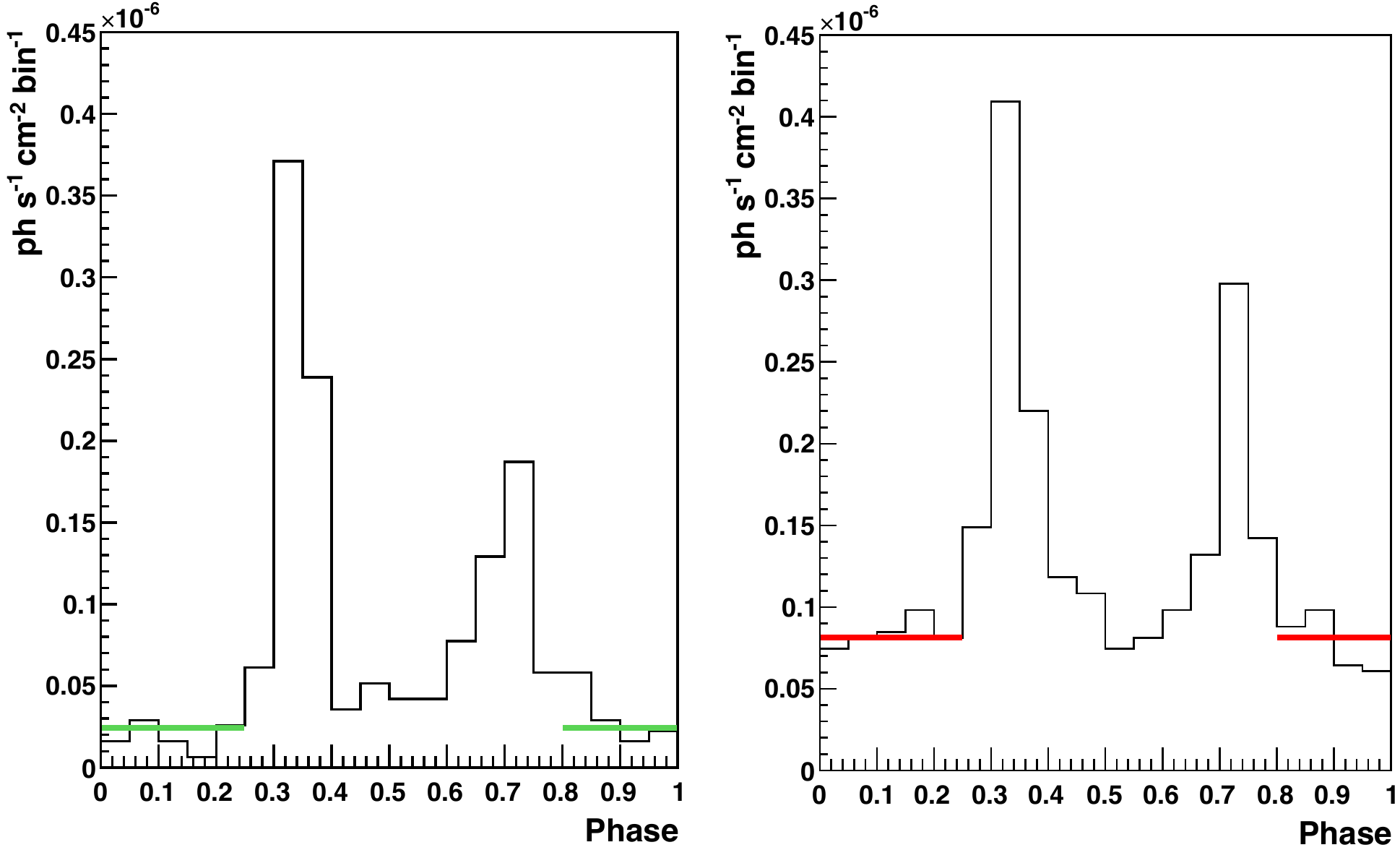}}
   \caption{Exposure corrected \object{Crab} pulse profiles.
   \textit{Left panel}: before the flares. \textit{Right panel}: during the flares. The horizontal green and red lines indicate the off-pulse averages.}
   \label{fig:pulse_fermi}
   \end{figure}

In Fig.~\ref{fig:pulse_fermi}, we compare the pulse profile measured before and during the flares. We can estimate the contribution of the nebula, by assuming that the pulsar emission is negligible in the phase ranges $\phi<0.25$ and $\phi>0.8$.
Before the flares, the average rate of the nebular emission is $(2.44 \pm 0.29) \times 10^{-8}~\mathrm{ph~cm^{-2}~s^{-1}~bin^{-1}_{phase}}$, whereas during the flare it is $(8.12 \pm 0.55) \times 10^{-8}~\mathrm{ph~cm^{-2}~s^{-1}~bin^{-1}_{phase}}$. After subtracting the contribution from the nebula, we compute the average rate of the pulsed emission: before the flare it is $(9.32 \pm 0.66) \times 10^{-8}~\mathrm{ph~cm^{-2}~s^{-1}~bin^{-1}_{phase}}$, whilst during the flares it is $(8.52 \pm 0.90) \times 10^{-8}~\mathrm{ph~cm^{-2}~s^{-1}~bin^{-1}_{phase}}$. During the flares, we conclude that  the flux from the nebula increased by a factor $3.33 \pm 0.46$. In contrast, the pulsar flux remained constant within the errors. This result confirms the preliminary claim of \cite{2010ATel.2879....1H}.

To study the high energy spectrum of the \object{Crab}, we exploit the maximum likelihood method for the dataset listed in Table~\ref{tab:intervals}. Following \citet{2010ApJ...708.1254A}, we model the spectral emission from the \object{Crab} using two power-law components for the nebula, which represent respectively the synchrotron and IC emissions, and a power-law with an exponential cutoff to describe the pulsar contribution. As the flare is not pulsed, we fix the parameter related to the pulsar emission to these of \citet{2010ApJ...708.1254A}, and assume that the IC contribution of the nebula has not changed during the flare. The only free parameters are the synchrotron contribution and the normalization of the Galactic emission. The synchrotron emission is found to increase from a flux of $(5.6 \pm 1.3)\times 10^{-7}\,\mathrm{ph~cm^{-2}~s^{-1}}$ to $(32.4 \pm 2.7)\times 10^{-7}\,\mathrm{ph~cm^{-2}~s^{-1}}$ in the 0.1-300~GeV band.

\section{Discussion}

To discriminate among the various models capable of reproducing the quasi-exponential turnover of the synchrotron emission of the nebula that peaks below the LAT energy window, we studied the Fermi data complemented by archival CGRO/COMPTEL data (0.75-30 MeV). A single power-law cannot reproduce the pre-flare data ($\chi^2 / d.o.f. \sim 44 / 15$). A power-law with a high energy exponential cutoff can instead reproduce the data ($\chi^2 / d.o.f. \sim 4 / 14$). To model the nebular synchrotron spectrum, we used the following function
\begin{equation}
\frac{\mathrm{d}N}{\mathrm{d}E} = N_0 \left( \frac{E}{1\,\mathrm{GeV}} \right)^{-\Gamma} \mathrm{exp}\left( - \frac{E}{E_{cutoff}} \right).
\end{equation}

\begin{figure}[t!]
   \resizebox{\hsize}{!}{\includegraphics[width=0.49\textwidth]{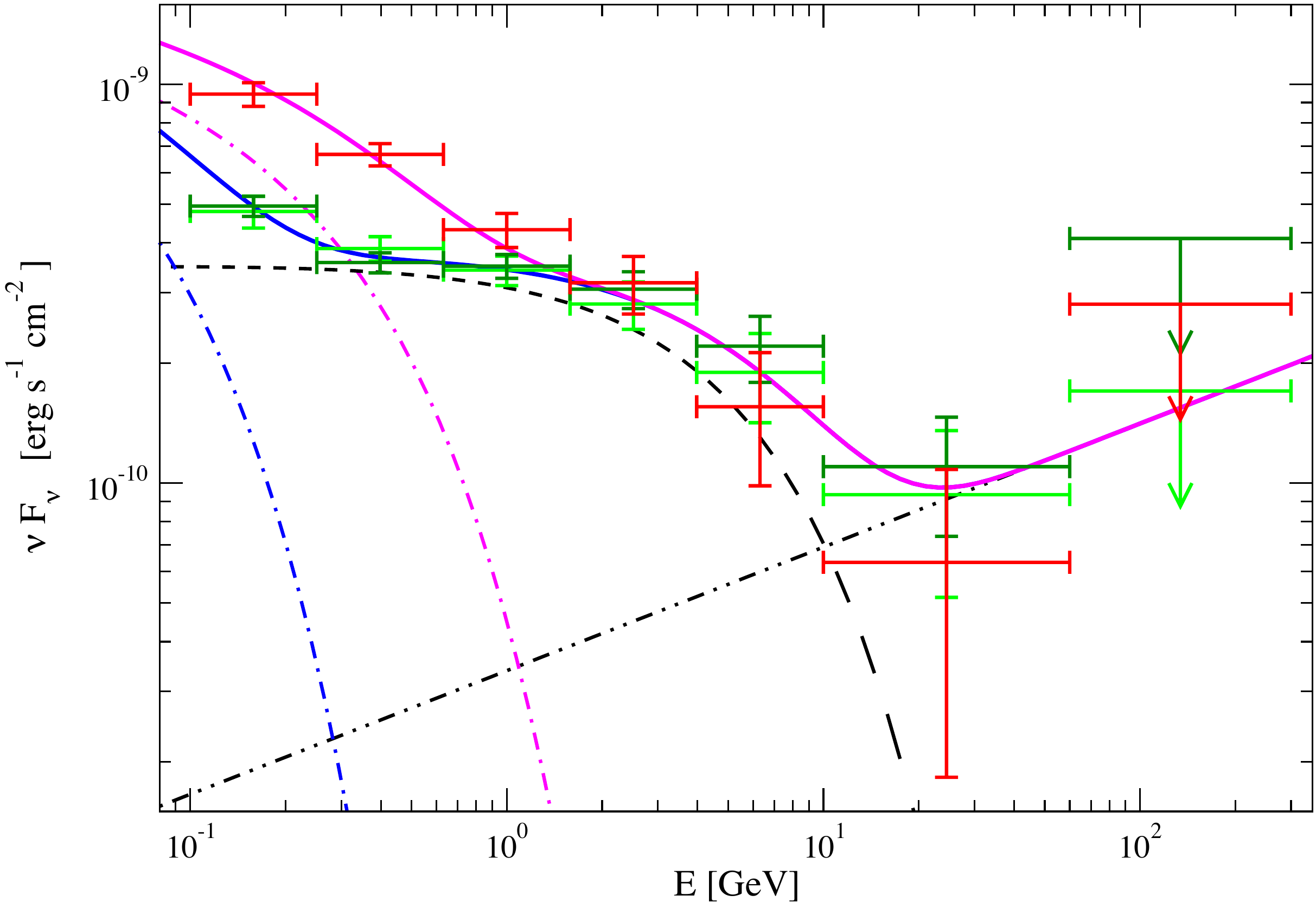}}
   \caption{\object{Crab} spectral energy distribution in the 100 MeV - 300 GeV energy range. The points with error bars are the Fermi detections before the flare (dark green), during the flare (red), and after the flare (light green). The black dashed line represents the contribution from the pulsar. The black dot-dot-dashed line represents the IC emission from the nebula. The blue and magenta dot-dashed and solid lines are the synchrotron nebula and the total emission before and during the flare, respectively. Arrows indicate the 95\% confidence flux limits.}
   \label{fig:sed}
   \end{figure}

The best-fit solution yields $\Gamma = -2.20 \pm 0.08$ and $N_0 = (4.3 \pm 1.9) \cdot 10^{-10}$ $\mathrm{ph~cm^{-2}~s^{-1}~MeV^{-1}}$. The difference between the quiescent and flaring spectra can be understood by considering two different extreme cases of either a constant power-law normalisation or a constant cutoff energy. In the former case, an increase in the energy cutoff of a factor of nearly 5 (from $77 \pm 15$ MeV to $367 \pm 45$ MeV) is needed (as illustrated in Fig.~\ref{fig:sed}). This increase is averaged over the whole flaring period, thus represents a lower limit, since in each single flare the cutoff energy might have been higher. In the latter case, the spectral variability can be explained by raising the continuum normalization by a factor of $\sim5$.

We note that the non-detection of any significant hard X-ray variability during the flare does not allow us to differentiate between the two possibilities as several electron populations are probably present in the nebula. As reported by \cite{1998nspt.conf..439A}, the COMPTEL data are characterized by a flattening of the spectrum that can be ascribed to the synchrotron emission of a separate electron population confined in compact regions such as wisps or knots. The luminosity of this component is less than 1\% of that of the whole nebula. 

The cutoff in the synchrotron spectrum occurs at a characteristic frequency $\nu_{\rm peak} \sim 4.2 \times 10^{6} \, B \, \gamma^{2}$. Provided that synchrotron radiation is the dominant mechanism through which particles channel their energy, the maximum electron Lorentz factor obtained by equating $t_{\rm sync}$ to $t_{\rm accel}$ is $\gamma_{\rm max} \propto (B \, \eta)^{-1/2}$, where $\eta \geq 1$ is the gyrofactor that characterizes the acceleration rate $\dot{\gamma}_{\rm accel} \equiv \gamma/t_{\rm accel}$ and $t_{\rm accel} = \eta \, E/q_{\rm e}\, B\,c$. This makes $\nu_{\rm peak}$ independent of $B$, leading to an electron synchrotron energy cutoff  $\approx 160 \eta^{-1}$~MeV \citep[see e.g.][]{2000NewA....5..377A}. A higher value may imply that the conditions in the accelerator differ from those in the emission region, e.g. there is a lower magnetic field in the former, or that the synchrotron gamma-rays are produced in a relativistically moving region, which produces a shift in the energy cutoff to higher energies by the corresponding Doppler factor $\delta$. In this scenario, a value $\delta \sim 367/160 \approx 2.3$ would be required to explain the energy cutoff obtained during the flaring episode.

On the other hand, magnetic fields at the level of between $\sim$~300~$\mu$G and $\sim$~2~mG are found in the synchrotron nebula and wisps, respectively \citep[see e.g.][]{2008ARA&A..46..127H}. Synchrotron radiation at $\sim$~1~GeV implies that $\gamma \sim 3-10 \times 10^{9}$ in the emitting regions. Taking $\delta \sim 2.3$, the comoving cooling timescale for those particles, taking an extreme value $B \sim$ 2 mG, is $\sim 0.3$~d. The corresponding observer timescale would then be similar to the decay time of the peaks present in the \textit{Fermi} lightcurve during the flaring period, $\lesssim 1$~d. In contrast, the flares could be related to an enhanced electron population, and the spectral variability could be obtained by raising the continuum normalization by a factor of $\sim 5$ or by adding a hard very high energy electron population (leading to a photon index $\Gamma<1$). In this case, a Doppler boosting may not be required, and the observed  duration of the flares could correspond to the synchrotron timescale of PeV electrons embedded in magnetic fields $\lesssim 1\,\mathrm{mG}$. 

The duration of the three short flares limits the size of the emitting region(s) to $\lesssim 10^{15}$ cm. The peak luminosity of these flares is higher/brighter than $10^{35}$ erg/s, i.e. $\geqslant 0.5$\,\textperthousand \,\,of the \object{Crab} spin-down luminosity, assuming an isotropic distribution. The distance between the emitting region and the pulsar can thus be constrained to be $\leqslant 6\times 10^{16}$ cm, i.e. not larger than 15\% of the size of the bright synchrotron torus observed by Chandra and HST, and probably consistent with the half-width of this torus. The emitting region could therefore be linked to the interaction zone between the jet and the torus, which is found to have brightened in the HST image obtained on 2 October \citep{2010ATel.2903....1C}. The three flares separated by two days could possibly be related to various emitting knots in this region. Alternatively, gamma-rays could be produced within the jet itself. However, if the emitting region were moving at relativistic speeds, the emission would be radiated within an angle $\sim 1/\delta$. For reasonable values of the jet inclination angle with respect to the line of sight \citep[see e.g.][]{2004ApJ...601..479N}, this scenario would make the flares difficult to detect.

To conclude, the flare relative short durations ($< 1$~day), their soft spectrum, and the analysis of the pulse profile in the 0.1-300\,GeV indicate that one or more compact portions ($\lesssim 10^{15}$~cm or $<$~0.1'') of the synchrotron nebula are responsible for the flares. In these region(s), freshly accelerated PeV electrons are rapidly cooling, causing the observed variability.

\begin{acknowledgements}
PB has been supported by the DLR grant 50 OG 1001.
\end{acknowledgements}

\bibliographystyle{aa}
\bibliography{15980}

\end{document}